\documentstyle[12pt,aaspp4]{article}

\lefthead{Halpern et al.}
\righthead{EUVE J0425.6--5714: A Newly Discovered AM Herculis Star}

\begin{document}

\title{EUVE J0425.6--5714: A Newly Discovered AM Herculis Star}

\author{J. P. Halpern and K. M. Leighly}
\affil{Columbia Astrophysics, Columbia University, 550 West 120th Street,
       \\New York, NY 10027}

\author{H. L. Marshall}
\affil{Eureka Scientific, Inc., 2452 Delmer St., Suite 100, Oakland, CA 94602}

\author{M. Eracleous\altaffilmark{1}}
\affil{Department of Astronomy, University of California, Berkeley, CA 94720}

\and

\author{T. Storchi-Bergmann\altaffilmark{2}}
\affil{Departamento de Astronomia, IF-UFRGS, CP 15051, CEP 91501-970,
       \\Porto Allegre, RS, Brasil}

\altaffiltext{1}{Hubble Fellow}

\altaffiltext{2}{Visiting Astronomer, Cerro Tololo Inter-American Observatory, 
National Optical Astronomy Observatories, which is operated by AURA, Inc.,
under a cooperative agreement with the National Science Foundation.} 

\begin{abstract}
We detected a new AM Her star serendipitously
in a 25~day observation with the {\it EUVE\/} satellite.
A coherent period of 85.82~min is present in the {\it EUVE\/} Deep Survey 
imager light curve of this source.
A spectroscopic optical identification is made with a 19th magnitude
blue star that has H and He emission lines, and broad cyclotron humps typical
of a magnetic cataclysmic variable.  A lower limit to the polar
magnetic field of 46~MG is estimated from the spacing of the cyclotron
harmonics.  EUVE J0425.6--5714 is also detected in archival
{\it ROSAT\/} HRI observations spanning two months, and its stable and highly 
structured light curve permits us to fit a coherent ephemeris linking the
{\it ROSAT} and {\it EUVE} data over a 1.3~yr gap.
The derived period is $85.82107 \pm 0.00020$~min,
and the ephemeris should be accurate to 0.1 cycles until the
year 2005.  A narrow but partial X-ray eclipse suggests that this object
belongs to the group of AM~Her stars whose viewing geometry is such that
the accretion stream periodically occults the soft X-ray emitting accretion
spot on the surface of the white dwarf.  A non-detection of hard
X-rays from {\it ASCA} observations that are contemporaneous
with the {\it ROSAT\/} HRI shows that the soft X-rays must dominate
by at least an order of magnitude, which is consistent with a
known trend among AM~Her stars with large magnetic field.

This object should not be confused with the Seyfert galaxy 1H~0419--577 
(= LB~1727), another X-ray/EUV source
which lies only $3.\!^{\prime}95$ away, and was the principal
target of these monitoring observations.
\end{abstract}


\section{Introduction}

AM Her stars, or polars, are the subclass of cataclysmic binaries
in which the white dwarf is highly magnetized, typically
$B_{\rm p} > 10$~MG, and Roche-lobe overflow
from a low-mass companion proceeds through an accretion stream
directly onto the magnetic pole(s) of the white dwarf without the
intermediary of an accretion disk.  The magnetic field in polars is also
strong enough to synchronize the rotation of the white dwarf
with the orbital period of the system.  Hard X-rays are produced by
bremsstrahlung from the hot gas which is heated by an accretion shock
above the surface of the white dwarf.  Soft X-ray/EUV radiation is
powered by reprocessing of the hard X-rays on the surface, and also by direct
heating of the subphotospheric layers by dense blobs raining down
from the accretion stream.  Most of the known AM Her stars were discovered 
because of their X-ray radiation.  For reviews, see Cropper (1990) and
Beuermann \& Burwitz (1995).

This is the third in a series of papers which demonstrate the value of
long {\it EUVE} observations for obtaining interesting timing results
on more than one source in the field (see Halpern, Martin, \& Marshall 1996;
Halpern \& Marshall 1996).

\section{EUVE Observation}

We began a long {\it EUVE\/} observation of the Seyfert galaxy 1H~0419--577
($z = 0.104$) on 1997 December 15.
It was immediately apparent in quick-look data that a {\it pair}
of sources of equal brightness was present near the target position.
Only $3.\!^{\prime}95$ apart, they are nevertheless well separated
in the {\it EUVE\/} Deep Survey imager.  This instrument is sensitive
in the range 65--190 eV, although interstellar absorption limits the
detected flux from extragalactic sources to energies greater than 100~eV.
Figure~1 shows the light curves of both sources,
where each point represents one satellite orbit.  In order to construct
these light curves, counts were extracted
from a circular aperture of radius $75^{\prime\prime}$ around each source.
Background was obtained from adjacent regions and subtracted, and the light
curves were corrected for variable dead time and ``Primbsching'' (lost counts
due to telemetry sharing).  The total exposure time is 651,405~s, and the
net counts are $44,403$ and $44,498$, respectively, from the Seyfert galaxy
1H~0419--577 and the new source denoted EUVE J0425.6--5714.  Table~1 contains
a log of this observation.
 
\begin{figure}
\plotone{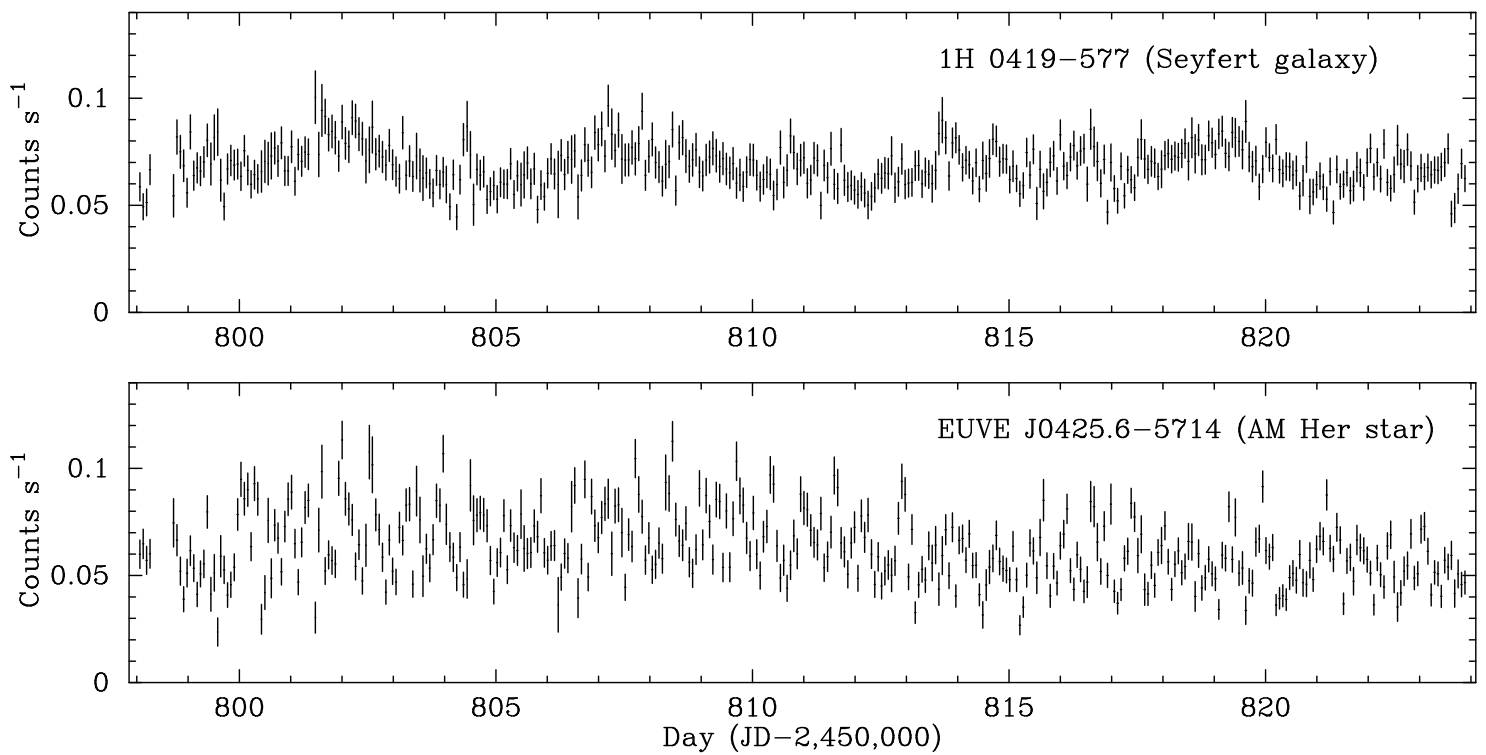}
\vskip -0.3 truein
\caption{Light curves of the two neighboring sources in the
{\it EUVE\/} Deep Survey imager. Each point corresponds to one satellite
orbit.  Background subtraction and all relevant exposure corrections have
been applied.   The fact that the two sources
do not show correlated variability assures us that the
periodic signal detected in EUVE J0425.6--5714 is real, and not the result of
some unknown systematic effect.
The short gap near the beginning of the observation is
due to an (unsuccessful) attempt to detect the afterglow of the gamma-ray
burst of 1997 Dec. 14, which occurred elsewhere on the sky
(Boer et al. 1997). \label{fig1}}
\end{figure}

While the Seyfert galaxy is only moderately variable, the new source shows
large-amplitude modulation which is the result of the beating between
the {\it EUVE\/} satellite orbit (94.6~min) and the intrinsic variability
of the source.  The latter was determined unambiguously
to be a $85.822 \pm 0.002$~min coherent period both from
epoch folding (chi-square) analysis, and from a discrete Fourier transform of
the individual photon arrival times.  The fact that the light curves
of the two sources
in Figure~1 do not show correlated variability assures us that the
periodic signal detected in EUVE J0425.6--5714 is real, and not the result of
some unknown systematic effect.
We interpret the 85.822~min period as both the orbital period and the spin 
period of a synchronously rotating magnetic CV (see \S 4).
We searched for additional periods as short as 10~s, but none were found.

\begin{deluxetable}{crrrrrrrrrrr}
\footnotesize
\tablecaption{X-ray Observations of EUVE 0425.6--5714. \label{tbl-1}}
\tablewidth{0pt}
\tablehead{
\colhead{} & \colhead{Date} & \colhead{Exposure Time}   & \colhead{Count Rate} 
\\
\colhead{Instrument} & \colhead{(UT)} & \colhead{(s)} & \colhead{(s$^{-1}$)} 
} 
\startdata
{\it ROSAT} PSPC & 1992 April 7               &     4094 &  $<0.0025$ \nl
{\it ROSAT} HRI  & 1994 May 15                &     2114 &   0.1420   \nl
{\it ROSAT} HRI  & 1994 Sep 19--23            &     6015 &   0.0387   \nl
{\it ROSAT} HRI  & 1996 Jun 30 -- Sep 1       &  171,841 &   0.0289   \nl
{\it ASCA} SIS   & 1996 July 22               &   24,370 &  $<0.0030$ \nl
{\it ASCA} SIS   & 1996 Aug 10                &   24,397 &  $<0.0033$ \nl
{\it EUVE} DS    & 1997 Dec 15 -- 1998 Jan 10 &  651,405 &   0.0683   \nl
\enddata
\end{deluxetable}

Figure~2 is the mean folded light curve of EUVE J0425.6--5714.
All phase bins are well covered,
despite the proximity of the source period and the
satellite orbital period, because the observation is very long.
Phase 0 is defined as the midpoint
of the narrow dip feature.  The partial reversal inside the dip
appears to be real, and is present throughout the entire 25 day observation.
In fact, the entire light curve appears to be stable over the duration of this
observation.  An ephemeris for mid-dip (phase 0) is

$T_{\rm dip}({\it EUVE}) = {\rm HJD}\,2450798.08606(60) + 
0.0595990(14)\times E$

\noindent
The numbers in parentheses are the uncertainties in the last digits.

\begin{figure}
\plotone{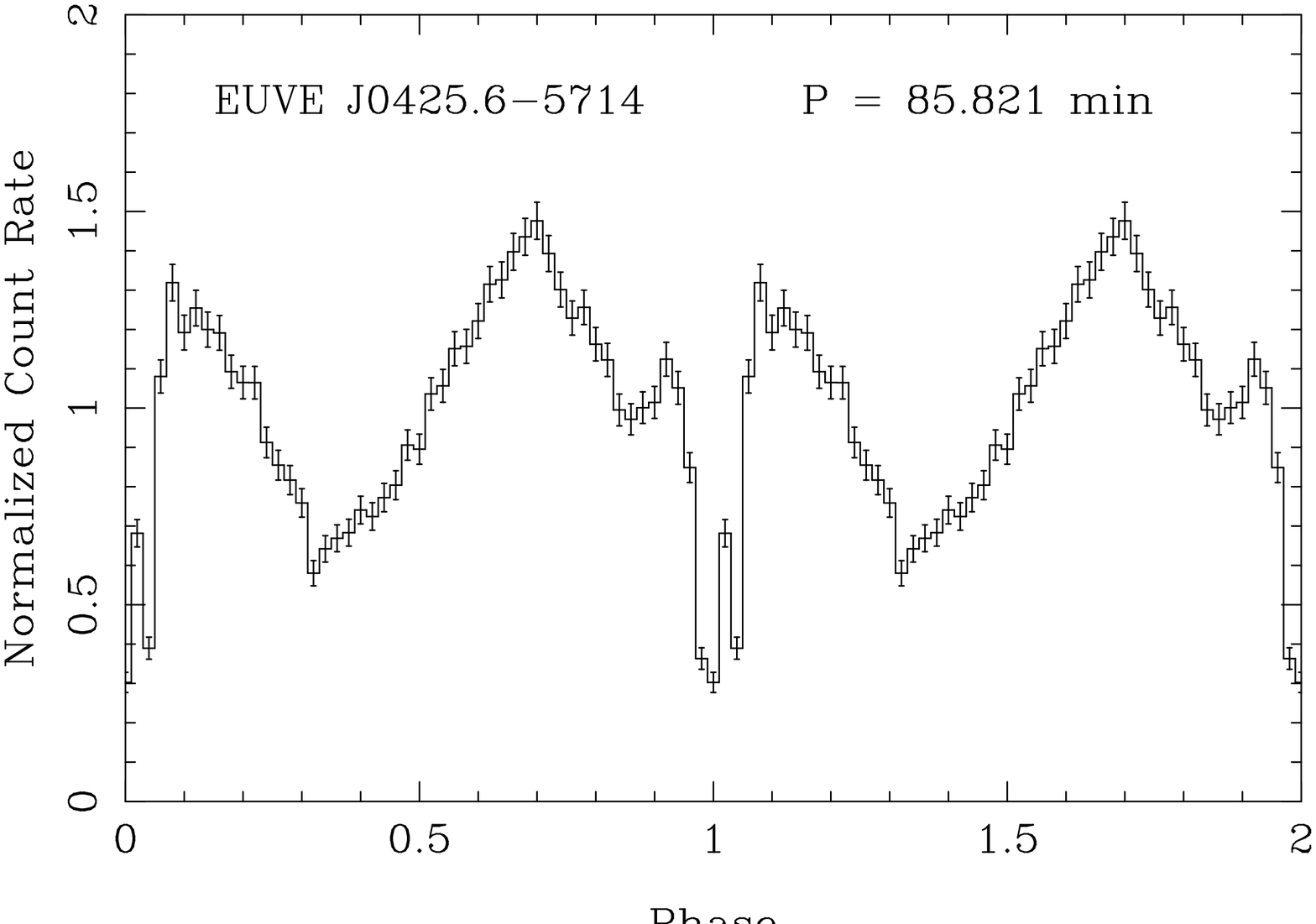}
\vskip 0.3 truein
\caption{Folded light curve of the AM~Her star EUVE J0425.6--5714
in the {\it EUVE\/} Deep Survey imager.  Background has been subtracted.
Phase~0 corresponds to HJD 2450798.08606, the epoch
of the ephemeris in \S 2. \label{fig2}}
\end{figure}

\section{ROSAT and ASCA Observations}

This field has been observed many times by {\it ROSAT\/}, also with
the Seyfert galaxy 1H~0419--577 as the intended target.
A log of the {\it ROSAT\/}
observations is given in Table~1.  An X-ray source at the position of
EUVE J0425.6--5714 is present in {\it ROSAT\/} HRI observations beginning
in 1994 May, when it may have been in its brightest X-ray state.
The HRI is sensitive in the range 0.2--2.0~keV.
This source was not detected in the only {\it ROSAT\/} PSPC
observation, obtained on 1992 April 7. Since the PSPC is 
several times more sensitive than the HRI at all energies,
the source must have been in an ``off'' state
at the time of the PSPC observation.
There are no reports of previous soft X-ray detections of
EUVE J0425.6--5714, and two {\it ASCA\/} pointings obtained contemporaneously
with the {\it ROSAT\/} HRI detection (see Table 1) show that it is apparently 
not a hard X-ray source in the 0.5--10~keV range,
even in the ``on'' state.  Because the {\it ASCA} 
image of the bright Seyfert galaxy has a complex, extended point-spread
function, the upper limits quoted in Table~1 were
derived with the help of ray-tracing software, written by Andy Ptak, which
we used to simulate the appearance of a second source at the location of
the star.

The HRI monitoring observations obtained in 1996
have sufficient exposure to produce a folded light curve.
Here we describe the analysis of EUVE J0425.6--5714
from these repeated HRI observations.
A total exposure time of 171,841~s was divided into approximately daily
observations of one or two satellite orbits each, spanning a two-month
period.  In addition, denser coverage was obtained for 5 days within
this interval.  From all these observations, a total of 4974 net counts were
extracted from the source after background subtraction.
The average flux was fairly steady throughout this period. 
Epoch folding of these data reveal a highly
significant signal at a period consistent with the {\it EUVE\/} measured
value.  The best fitting period in the HRI is $85.8213 \pm 0.0016$~min.
The HRI folded light curve, shown in Figure~3, is nearly identical
to that of {\it EUVE\/}, including the timing of the narrow dip relative
to the broad modulation. An ephemeris for mid-dip (phase 0) in the HRI is

$T_{\rm dip}({\rm HRI}) = {\rm HJD}\,2450264.5651(12) + 
0.0595981(11)\times E$

\begin{figure}
\plotone{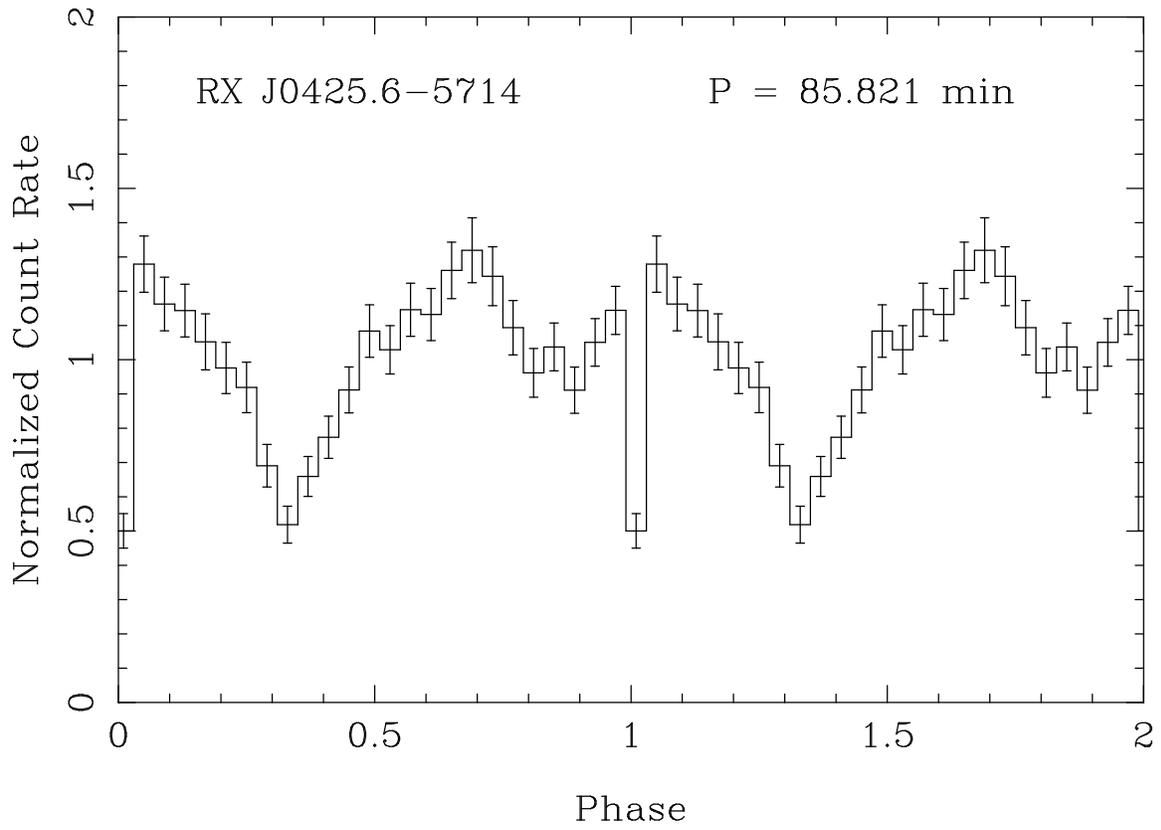}
\vskip 0.3 truein
\caption{Folded light curve of the AM~Her star EUVE J0425.6--5714
in the {\it ROSAT\/} HRI observations of 1996.  Background has been
subtracted.  Phase~0 correspond to HJD 2450264.5651, the epoch of the ephemeris 
in \S 3. \label{fig3}}
\end{figure}

As a result of the apparent stability of the light curve,
the sharpness of the dip, and the long time span of
the observations, both of the above ephemerides are
precise enough to count cycles over the 1.5~yr interval between
the epochs $T_0({\rm HRI})$ and $T_0({\it EUVE})$.
The HRI ephemeris predicts that $8951.98 \pm 0.15$ cycles elapsed between
$T_0({\rm HRI})$ and $T_0({\it EUVE})$, while the {\it EUVE\/} ephemeris
predicts that $8951.84 \pm 0.21$ cycles elapsed.  Therefore, we conclude
that there is a unique cycle count of 8952 between
the epochs.  The resulting coherent ephemeris is  

$T_{\rm dip} = {\rm HJD}\,2450264.5651(12) + 0.05959796(13)\times E$

\noindent
The resulting best period, $85.82107 \pm 0.00020$~min, is consistent with the
values measured individually by {\it ROSAT\/} and {\it EUVE\/}.  This
joint ephemeris is expected to be accurate to 0.1 cycles until the
year 2005.

\section{Optical Identification and Spectroscopy}

Armed with a precise X-ray position, which was confirmed by relative astrometry
with respect to the known AGN, it was a simple matter to identify
EUVE J0425.6--5714 with a 19th magnitude blue star on the ESO
Sky Survey plates.  This star is listed in the USNO A1.0 astrometric
catalogue (Monet et al. 1996)
at J2000 coordinates $4^{\rm h}25^{\rm m}38.\!^{\rm s}65$;
$-57^{\circ}14^{\prime}36.\!^{\prime\prime}5$, with approximate magnitudes
$B = 19.1, R = 18.7$.  A finding chart is given in Figure~4.  The
Seyfert galaxy 1H~0419--577 is also marked on this chart, at J2000 coordinates
$4^{\rm h}26^{\rm m}0.\!^{\rm s}76$;
$-57^{\circ}12^{\prime}1.\!^{\prime\prime}6$.
We note that the position of this almost stellar Seyfert galaxy
is consistent with that of the blue ``star'' LB~1727, given as
(1950) $4^{\rm h}25.\!^{\rm m}0$; $-57^{\circ}19^{\prime}$ by
Luyten \& Anderson (1958).  However, there are several references to
an erroneous position for this Seyfert galaxy
in the literature.  The finding chart in Shara et al. (1993)
for the {\it ROSAT\/}
Wide Field Camera catalogue points to the wrong object, as does
Shara et al. (1997) for the {\it EUVE\/} catalogue.
As a result, an incorrect position was
measured by Veron-Cetty \& Veron (1996), and was adopted by
several of the meta-catalogues that are in wide use today. 

\begin{figure}
\plotone{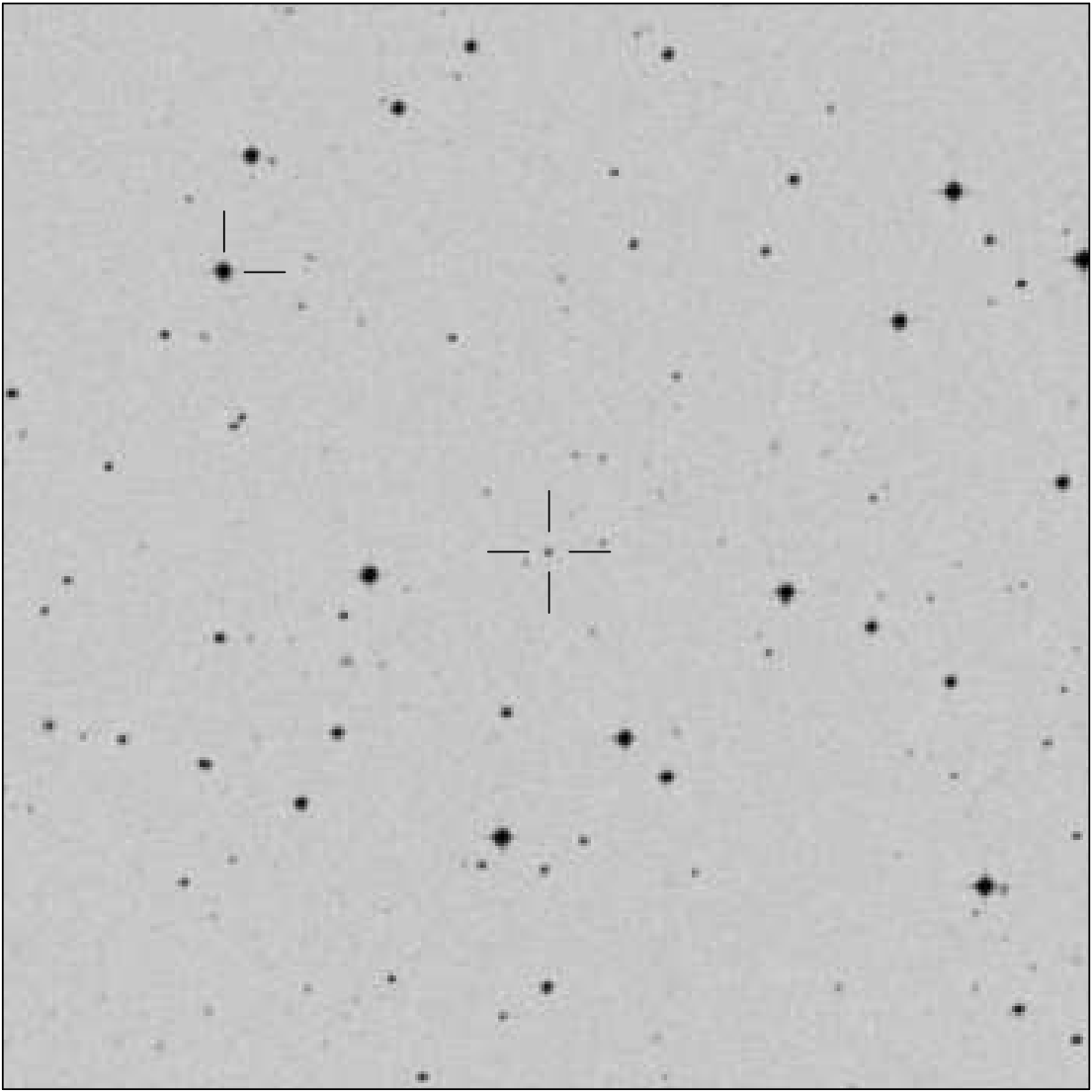}
\vskip -0.3 truein
\caption{A finding chart that you can trust for 
the AM~Her star EUVE J0425.6--5714 at J2000 coordinates
$4^{\rm h}25^{\rm m}38.\!^{\rm s}65$;
$-57^{\circ}14^{\prime}36.\!^{\prime\prime}5$ (center),
and the Seyfert galaxy 1H~0419--577
(= LB~1727) at J2000 coordinates
$4^{\rm h}26^{\rm m}0.\!^{\rm s}76$;
$-57^{\circ}12^{\prime}1.\!^{\prime\prime}6$ (upper left).
The field is $10^{\prime}\times 10^{\prime}$ from
a digitized SERC Southern Sky Survey IIIa-J plate.
North is up, East is left, black is white.
\label{fig4}}
\end{figure}

We obtained optical spectra of the 19th magnitude optical counterpart of
EUVE J0425.6--5714 on the CTIO 1.5m telescope on 1998 January 3 and 4,
simultaneously with the ongoing {\it EUVE\/} observation.
Figure~5 shows a summed
spectrum amounting to 1~hr of exposure.  Its continuum flux corresponds
to $B = 19.3$.  When accounting for slit losses, this is probably
consistent with its appearance on the ESO Sky Survey, and with the 
USNO A1.0 catalogue. Strong emission lines of H, He~I, and He~II~$\lambda$4686
are seen, together with the broad cyclotron humps that are
characteristic of AM Her type cataclysmic binaries.  Cyclotron peaks
are located at approximately 4350 \AA\ and 5350 \AA.  Their separation
of 4300 cm$^{-1}$ corresponds to a lower limit on the polar magnetic field
of 46 MG according to equation (4) of Cropper (1988).  An actual measurement
of the magnetic field and the temperature/viewing parameter
$T\,{\rm sin}^2\theta$ would require time-resolved spectra with broader
wavelength coverage to identify and accurately measure several more harmonics.

\begin{figure}
\plotone{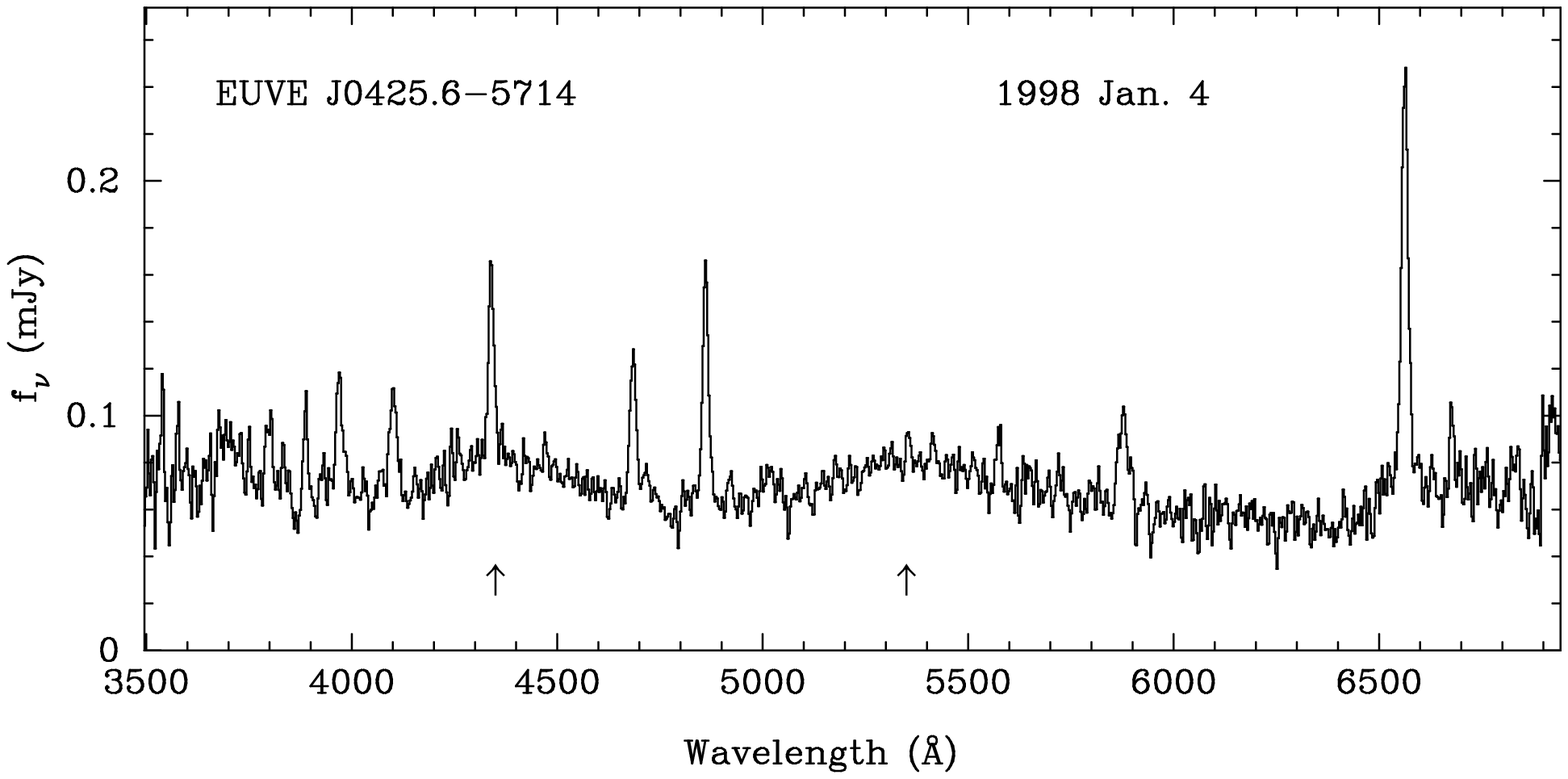}
\caption{One hour of exposure on the AM Her star
EUVE J0425.6--5714 obtained on the CTIO 1.5m telescope.  All the
telltale signatures are here (strong He~II~$\lambda$4686, cyclotron humps).
Arrows indicate the locations of cyclotron harmonic peaks.
\label{fig5}}
\end{figure}

\section{Interpretations and Conclusions}

It is difficult to derive quantitative information about luminosity
and distance to this system from existing data.  We did not detect the
secondary star in the optical spectrum; for this short orbital
period it is expected to be a late M dwarf.  Although the {\it EUVE\/}
and {\it ROSAT\/} light curves are of high quality, they lack any spectral 
information from which temperatures and fluxes can be measured.
(The source is too faint for a useful spectrum to be extracted from
the {\it EUVE\/} short wavelength spectrometer).  However,
based on the similarity of the {\it EUVE\/} and {\it ROSAT\/} light curves,
we conclude that both of these represent soft blackbody emission
coming from the heated accretion spot(s) on the white dwarf,
and we hypothesize
that the accretion rate may have been similar during both epochs.
Under these assumptions,
the ratio of the {\it EUVE\/} and {\it ROSAT\/} count rates can
be interpreted in terms of an allowed range of blackbody temperature
and intervening column density.  We find that the range of parameters that
is consistent with both count rates is $21 < {\rm k}T < 30$~eV and
$0 < N_{\rm H} < 3 \times 10^{19}$~cm$^{-2}$, with k$T$ strongly anticorrelated
with $N_{\rm H}$.  The lower bound on the temperature is actually set
by the additional requirement that the Rayleigh-Jeans tail of the 
blackbody not exceed the observed $U$-band flux obtained from Figure~5.
The actual effective temperature of a real stellar atmosphere could,
however, be lower than the blackbody estimate due to non-grey opacity.
The bolometric flux of these fitted blackbody models ranges from
$(1.4-6.8) \times 10^{-12}$ ergs~cm$^{-2}$~s$^{-1}$.

With respect to hard X-rays, there is a dearth of evidence for any
from EUVE J0425.6--5714.
The two {\it ASCA} non-detections listed in Table~1 correspond
to a combined 3$\sigma$ upper limit on the hard X-ray flux of
$3.78 \times 10^{-14}$ ergs~cm$^{-2}$~s$^{-1}$ in the 0.5--2.0~keV band,
assuming a 10~keV thermal bremsstrahlung spectrum. Since these observations
were contemporaneous with the {\it ROSAT} HRI monitoring, we can use the
{\it ASCA} non-detections to place an upper limit on the fraction of
the HRI flux that
can be thermal bremsstrahlung. A flux of $3.78 \times 10^{-14}$
ergs~cm$^{-2}$~s$^{-1}$ in the 0.5--2.0~keV band 
would produce $\approx 0.0018$
counts~s$^{-1}$ in the HRI.  Since the actual HRI count rate is 0.029
counts~s$^{-1}$, a factor of 16 larger, we can safely conclude that
EUVE J0425.6--5714 is one of those AM~Her stars whose soft X-ray luminosity is
at least an order of magnitude larger than its hard X-ray luminosity.
Hard X-ray emission in AM Her stars is
attributed to bremmstrahlung from an accretion shock above the surface of the
white dwarf.   Approximately half of this energy should be reprocessed into
themal soft X-rays on the surface.  But several other AM~Her stars are 
known to be very soft sources without significant hard X-ray flux
(e.g., deMartino et al. 1998).  Beuermann \& Burwitz (1995)
showed that the ratio of bremsstrahlung flux to blackbody flux in the
{\it ROSAT\/} band is less than 0.05 for all systems in which $B>30$~MG.
Our lower limit of 46~MG from the cyclotron features is consistent
with this correlation.  When soft X-ray emission dominates over hard 
X-rays, it is attributed either to the suppression of bremsstrahlung
by the more efficient cyclotron cooling which occurs in a high magnetic field
(Woelk \& Beuermann 1996), or to accretion in the form
of dense blobs which deposit their energy below the photosphere of
the white dwarf (Kuijpers \& Pringle 1982). 

The narrow partial eclipse is most likely caused
by cold gas in the accretion stream that occults the hot spot.
The light curves of several AM~Her stars are interpreted in this
way (Patterson, Williams, \& Hiltner 1981;
Greiner, Remillard, \& Motch 1998 and references therein).
We note that the dip is approximately twice as broad in {\it EUVE\/}
as in {\it ROSAT\/}.  This favors an interpretation as photoelectric
absorption, whose angular extent can vary either as a function of
observed X-ray energy, with harder X-rays being less vulnerable to
absorption in the tenuous outskirts of the accretion stream, or as
a function of accretion rate.  In contrast, occultation by
the secondary star is more likely to produce a total eclipse that is
independent of energy, since both hard and soft X-rays
come from a relatively compact region near the surface of the white
dwarf.  The narrow reversal in the {\it EUVE} dip also argues against
eclipse by the secondary, even though we don't know exactly what
{\it does} cause it.  A similar feature appears to be present in
the {\it EUVE\/} light curve of QS~Tel (Rosen et al. 1996).

It is difficult to determine whether the light curve signifies
accretion onto one or both magnetic poles.  The waveform might be
described as double peaked with a separation of about 0.4 in phase,
indicating emission from two poles that are not diametrically opposite.
Alternatively, if the entire region between phase 0.7 and 0.1 is
affected by photoelectric absorption or scattering,
then the underlying light curve
resembles a broad, single peaked cosine function that is characteristic
of a single bright spot on a rotating star.  Several other AM~Her stars
have broad dips which precede an accretion-stream eclipse; their
EUV light curves have been modelled by Warren, Sirk, \&
Vallerga (1995), and Sirk \& Howell (1998).  These studies conclude
that either Compton scattering or absorption by ionized matter in the
accretion column just above the accretion spot can explain broad dips
which occur over phases which are not susceptible to occultation
by the part of the accretion stream that is far from the white dwarf.
Such models would seem to be consistent with the light curve of
EUVE J0425.6--5714, with a geometry such that the single accretion spot
is never completely occulted by the limb of the white dwarf.
Further progress in interpreting the accretion and viewing geometries of
this star will have to await spectrally resolved soft X-ray observations,
as well as phase-resolved optical spectroscopy and polarimetry.  The ephemeris
already obtained from existing X-ray data should permit absolute phasing
of any optical observations obtained over the next few years.

Results on the Seyfert galaxy 1H~0419--577 will be presented in a separate
paper.

\acknowledgments

We thank Joe Patterson for helpful discussions.
M. E. acknowledges support from Hubble fellowship grant
HF-01068.01-94A from the Space Telescope Science Institute, which is
oparated for NASA by the Association of Universities for Research in
Astronomy, Inc., under contract NAS~5-26255.


\clearpage

\begin{thebibliography}{}
\bibitem[Beuermann \& Burwitz 1995]{beu82} Beuermann, K., \& Burwitz, V. 1995,
    ASP Conf. Ser. 85, 99
\bibitem[Boer et al. 1997]{boe97} Boer, M., Roberts, B. A., Malina, R.,
    Feroci, M., Piro, L., \& Hurley, K. 1997, IAU Circ., 6795
\bibitem[Cropper 1988]{cro88} Cropper, M. 1988, MNRAS, 236, 29P
\bibitem[Cropper 1990]{cro90} Cropper, M. 1990, Space Sci. Rev., 54, 195
\bibitem[de Martino et al. 1998]{dem98} de Martino, D. et al. 1998,
    A\&A, 332, 904
\bibitem[Greiner, Remillard, \& Motch 1998]{gre98}
    Greiner, J., Remillard, R. A., \& Motch, C. 1998, A\&A, 336, 191
\bibitem[Halpern \& Marshall 1996]{hal96}
    Halpern, J. P., \& Marshall, H. L. 1996, ApJ, 464, 760
\bibitem[Halpern, Martin, \& Marshall 1996]{hmm96}
    Halpern, J. P., Martin, C., \& Marshall, H. L. 1996, ApJ, 462, 908
\bibitem[Kuijpers \& Primgle 1982]{kui82} Kuijpers, J., \& Pringle, J.
    1982, A\&A, 114, L4
\bibitem[Luyten \& Anderson 1958]{luy58}  Luyten, W. J., \& Anderson, J. H.
    1958, Minnesota University Observatory
\bibitem[Monet et al. 1996]{mon96} Monet, D. et al. 1996, USNO--SA1.0
    (Washington, DC: US Naval Observatory)
\bibitem[Patterson, Williams, \& Hiltner 1981]{pat81}
    Patterson, J., Williams, G., \& Hiltner, W. A. 1981, ApJ, 245, 618
\bibitem[Rosen et al. 1996]{ros96} Rosen, S. R., et al. 1996, MNRAS, 280, 1121
\bibitem[Shara et al. 1997]{sha97} Shara, M. M., Bergeron, L. E., Christian,
     C. A., Craig, N., \& Bowyer, S. 1997, \pasp, 109, 998
\bibitem[Shara et al. 1993]{sha93} Shara, M. M., Shara, D. J., \& McLean, B.
    1993, \pasp, 105, 387
\bibitem[Sirk & Howell 1998]{sir98} Sirk, M. M., \& Howell, S. B. 1998,
    astro-ph/9805182
\bibitem[Veron-Cetty \& Veron 1996]{Ver96} Veron-Cetty, M.-P., \& Veron, P.
     1996, A\&AS, 115, 97
\bibitem[Warren, Sirk, \& Vallerga 1995]{war95} Warren, J. K., Sirk, M. M.,
     \& Vallerga, J. V. 1995, ApJ, 445, 909
\bibitem[Woelk \& Beuermann 1996]{woe96}
     Woelk, U., \& Beuermann, K. 1996, A\&A, 306, 232
\end{thebibliography}
\end{document}